\documentclass[rmp,aps,onecolumn,nofootinbib]{revtex4}

\usepackage{amsmath}
\usepackage{tabularx}
\usepackage{palatino} 
\usepackage[]{graphicx}
\usepackage[]{setspace} 

\usepackage{epsfig}
\usepackage{epstopdf}

\newcommand{\Dkl}{D_{\mbox{\tiny KL}}}

\begin{document}
\begin{spacing}{1.5}

\title{Notes on Kullback-Leibler Divergence and Likelihood Theory}
\date{\today, version 1.01}
\author{Jonathon Shlens} 
\email{jonathon.shlens@gmail.com}
\affiliation{
Google Research \\
Mountain View, CA 94043}
\begin{abstract}
The Kullback-Leibler (KL) divergence is a fundamental equation of information theory that quantifies the proximity of two probability distributions. Although difficult to understand by examining the equation, an  intuition and understanding of the KL divergence arises from its intimate relationship with likelihood theory. We discuss how KL divergence arises from likelihood theory in an attempt to provide some intuition and reserve a rigorous (but rather simple) derivation for the appendix. Finally, we comment on recent applications of KL divergence in the neural coding literature and highlight its natural application. 
\end{abstract}
\maketitle

The Kullback-Leibler (KL) divergence is a measure in statistics \cite{Cover1991} that quantifies in bits how close a probability distribution $p = \{p_i\}$ is to a model (or candidate) distribution $q = \{q_i\}$,
\begin{equation}
\Dkl (p\;||\;q) = \sum_{i} p_i \log_2 \frac{p_i}{q_i}
\label{eq:kl}
\end{equation}
$\Dkl$ is non-negative ($\geq 0$), not symmetric in $p$ and $q$, zero if the distributions match exactly and can potentially equal infinity. A common technical interpretation -- although bereft of intuition -- is that the KL divergence is the ``coding penalty'' associated with selecting a distribution $q$ to approximate the true distribution $p$ \cite{Cover1991}.
\addtolength{\parskip}{\baselineskip}  

An intuitive understanding, however, arises from likelihood theory - the probability that one observes a set of data given that a particular model were true \cite{Duda2001}. Pretend we perform an experiment to measure a discrete, random variable - such as rolling a dice many times (or in neuroscience, the simultaneous binned firing patterns of multiple neurons). If we perform a long experiment and make $n$ measurements, we can count the number of times we observe each face of the die (or similarly, each firing pattern of neurons), a histogram $c = \{c_i\}$, where $n = \sum_i c_i$. This histogram measures the relative frequency of each face of the die (or, each type of firing pattern). If this experiment lasts forever, the normalized histogram counts $\frac{c_i}{n}$ reflect an underlying distribution $p_i = \frac{c_i}{n}$. Pretend we have a candidate model for die (or firing patterns), the distribution $q$. What is the probability of observing the histogram counts $c$ if the model $q$ actually generated the observations? This probability is given by the {\it multinomial likelihood} \cite{Duda2001},
$$L \propto \prod_i q_{i}^{c_i}$$
To gain some intuition, imagine that we performed $n=1$ measurements - in this case, the likelihood would be the $q_i$ attributed to the single observed firing pattern. The likelihood $L$ shrinks mutiplicatively as we perform more measurements (or $n$ grows). Ideally, we want the probability to be invariant to the number of measurements - this is given by the average likelihood $\bar{L} = L^{\frac{1}{n}}$, a number between 0 and 1. Matching intuition, as we perform more measurements, if $\frac{c_i}{n} \rightarrow q_i$, then the average likelihood would be perfect, or $\bar{L}\rightarrow 1$. Conversely, as $\frac{c_i}{n}$ diverges from the model $q_i$, the average likelihood $\bar{L}$ decreases, approaching zero. The link between likelihood and the KL divergence arises from the fact that if we perform an infinite number of measurements (see Appendix; \textcite{Shlens2006a}; Section 12.1 of \textcite{Cover1991}), 
\begin{equation}
\Dkl(p\;||\;q) = -\log_2 \bar{L}
\label{eq:kl-likelihood}
\end{equation}
Thus, if the distributions $p$ and $q$ are identical, $\bar{L} = 1$ and $\Dkl = 0$ (or if $\bar{L}=0$, $\Dkl = \infty$). The central intuition is that the KL divergence effectively measures the average likelihood of observing (infinite) data with the distribution $p$ if the particular model $q$ actually generated the data.

The KL divergence has many applications and is a foundation of information theory and statistics \cite{Cover1991}. For example, one can ask how similar a joint distribution $p(x,y)$ is to the product of its marginals $p(x)p(y)$ - this is the {\it mutual information}, a general measure of statistical dependence between two random variables \cite{Cover1991},
\begin{equation}
I(X; Y) = \sum_{x,y} p(x,y) \log_2 \frac{p(x,y)}{p(x)p(y)}
\label{eq:mutual-information}
\end{equation}
The mutual information is zero if and only if the two random variables $X$ and $Y$ are statistically independent. In addition to its role in mutual information, the KL divergence has been applied extensively in the neural coding literature, most recently to quantify the effects of conditional dependence between neurons~\cite{Schneidman2003b, Latham2005, Amari2006} and to measure how well higher order correlations can be approximated by lower order structure~\cite{Schneidman2006, Shlens2006a}.

\addtolength{\parskip}{-\baselineskip}  
\appendix
\section{Derivation}
In this appendix we prove that the relationship asserted in Equation~\ref{eq:kl-likelihood}. This derivation basically involves three main ideas: the application of Stirling's approximation, playing around with some algebra and recognizing an implicit probability distribution. First, we begin with some key some definitions.
\addtolength{\parskip}{\baselineskip}  

{\bf Multinomial likelihood}. \;\; The multinomial likelihood expresses the probability of observing a histogram, $c = \{c_i\}$ given that a particular model $q = \{q_i\}$ is true.
\begin{equation}
L(c|q) = \frac{n!}{\prod_i c_i !} \prod_i q_i^{c_i}
\label{eq:likelihood}
\end{equation}
The term in front $\frac{n}{\prod_i c_i !}$ is a normalization constant that counts the number of combinations which could give rise to the particular histogram. Note that $n = \sum_i c_i$ is the total number of measurements. 

{\bf Stirling's approximation}. \;\; Stirling's approximation, $\log n! \simeq n \log n - n$,  is a numerical approximation useful for large factorials that often appear in combinatorics. This approximation becomes quite good for $n > O(100)$.

We now begin the derivation by remembering that independent observations constituting a histogram are multiplied together to recover the joint probability of all measurements. Thus, an invariant likelihood across histogram counts is the geometric mean of the multinomial likelihood $L(c|q)^\frac{1}{n}$. We term this quantity the average multinomial likelihood, or {\it average likelihood} for short. We start by defining the average log-likelihood as
\begin{eqnarray*}
\bar{L} & \equiv & \log \; L(c|q)^\frac{1}{n}.
\end{eqnarray*}
Plugging in Equation~\ref{eq:likelihood} and a little algebra later,
\begin{eqnarray*}
\bar{L} & = & \frac{1}{n} \log \frac{n!}{\prod_i c_i !} \prod_i q_i^{c_i}\\
             & = & \frac{1}{n} \log n! - \frac{1}{n} \sum_i \log c_i ! + \sum_i \frac{c_i}{n} \log q_i
\end{eqnarray*}
We now plug in Stirling's approximation to simplify
\begin{eqnarray*}
\bar{L} & = & \frac{1}{n} \left( n\log n - n\right) - \frac{1}{n} \sum_i (c_i \log c_i - c_i) + \sum_i \frac{c_i}{n} \log q_i \\
& = & \log n - \sum_i \frac{c_i}{n} \log c_i + \sum_i \frac{c_i}{n} \log q_i
\end{eqnarray*}
Finally, rearranging terms highlights an implicit probability distribution.
\begin{eqnarray*}
\bar{L} & = & \sum_i \frac{c_i}{n}\log n - \sum_i \frac{c_i}{n} \log c_i + \sum_i \frac{c_i}{n} \log q_i \\
& = & -\sum_i \frac{c_i}{n}\log \frac{c_i}{n} + \sum_i \frac{c_i}{n} \log q_i
\end{eqnarray*}
In the limit of $n\rightarrow\infty$, the normalized histogram can be viewed as a probability distribution $p_i \equiv \frac{c_i}{n} $ and substituted accordingly.
\begin{eqnarray*}
\bar{L} & = & -\sum_i p_i \log p_i + \sum_i p_i \log q_i \\
& = & -\Dkl( p \; || \; q )
\end{eqnarray*}
where we now recognize the KL divergence (Equation~\ref{eq:kl}). The results can be summarized as
\begin{equation}
\Dkl (p \; || \; q ) = \lim_{n\rightarrow\infty} - \frac{1}{n}\log \; L(c|q)
\end{equation}
or the KL divergence is negative logarithm of the average multinomial log-likelihood.

A closer look at this derivation reveals that the normalization constant in front of Equation \ref{eq:likelihood} directly results in the term $-\sum_i p_i \log p_i$, which is the {\it entropy} of the distribution. Thus, it is possible to derive the entropy of a distribution from purely combinatorial notions~\cite{Jaynes2003}.


\addtolength{\parskip}{-\baselineskip}  
\bibliographystyle{apsrev}
\bibliography{kl}

\end{spacing}
\end{document}